\documentclass[journal,draftcls,onecolumn,12pt,twoside]{IEEEtran} 

\ifCLASSINFOpdf
\else
\fi
\usepackage{cite}
\usepackage{amsmath}
\usepackage{amssymb}
\usepackage{amsfonts}
\usepackage{psfrag}
\usepackage{epsfig} 
\usepackage{algorithm}
\usepackage{algorithmic}
\usepackage{color}

\newcommand{\B}[1]{\boldsymbol{#1}}
\DeclareMathOperator*{\argmin}{argmin}
\DeclareMathOperator*{\argmax}{argmax}

\begin{document}
%
\title{Massive MIMO Precoding and Spectral Shaping with Low Resolution {Phase-only} DACs and Active Constellation
Extension} 

\author{\IEEEauthorblockN{Amine Mezghani,~\IEEEmembership{Member,~IEEE} and Robert W. Heath Jr.,~\IEEEmembership{Fellow,~IEEE}} \\
\thanks{A.~Mezghani is with the ECE Department at the University of Manitoba, Winnipeg, MB, Canada (emails: \{Amine Mezghani\}@umanitoba.ca). R.~W.~Heath is with the North Carolina State University (email: rwheathjr@ncsu.edu). This work was supported by the US National Science Foundation (NSF), Grant No. ECCS-1711702 and CNS-1731658, and the Natural Sciences and Engineering Research Council of Canada (NSERC).}
}
\maketitle

\begin{abstract}
Nonlinear precoding and pulse shaping are jointly considered in multi-user massive multiple-input multiple-output (MIMO) systems with low resolution D/A-converters (DACs) in terms of algorithmic approach as well as large system performance. Two design criteria are investigated: the mean {squared} error (MSE) with active constellation
extension (ACE) and the symbol error rate (SER). Both formulations are solved based on a modified version of the generalized approximate message passing (GAMP) algorithm. Furthermore, theoretical performance results are derived based on the state evolution analysis of the GAMP algorithm. The MSE based technique is extended to jointly perform over-the-air (OTA) spectral shaping and precoding for frequency-selective channels, in which the spectral performance is characterized at the transmitter and at the receiver.
Simulation and analytical results demonstrate that the MSE based approach yields the same performance as the SER based formulation in terms of uncoded SER. 
The analytical results provide good performance predictions up to medium SNR. 
Substantial improvements in detection, as well as spectral performance, are obtained from the proposed combined pulse shaping and precoding approach compared to standard linear methods. 


\end{abstract}


%
\IEEEpeerreviewmaketitle

\section{Introduction}  
Multi-user precoding based on low resolution digital-to-analog converters (DACs) is an attractive approach for hardware-efficient and all-digital massive multiple-input multiple-output (MIMO) and millimeter-wave (mmWave) communications systems. It is desirable in terms of power efficiency and cost-performance tradeoff, to run the power amplifiers in saturated or switched-mode with constant envelope signals and low resolution phase-only quantizers, even down to one-bit. In fact, this enormous simplification of RF communications hardware can be compensated for by massive sampling in space and/or by advanced precoding algorithms \cite{Swindlehurst_2017}. The required precoders for multi-user separation are designed to enable constant envelope transmission and to operate under low resolution phase-only DACs \cite{Jacobsson_2017_2,Swindlehurst_2018}. This approach has been proposed
as a means to reduce power consumption, complexity, and cost of the RF frontend, analogously to the use of low resolution analog-to-digital converters (ADCs) at the receiver side as in prior work  \cite{nossek,ivrlac,ivrlac2006,mezghani2007,mezghaniisit2007,mezghani_ICASSP2008,mezghaniisit2008,Mo_2016,Zymnis_2010,mezghani_itg_2010,Yongzhiuplink,mollen2016performance,juncil2015near,jacobsson2015one,hea14,ning2015mixed}.

Different precoding algorithms have been proposed in MIMO systems with low resolution DACs. Besides the linear precoding techniques that have been considered in  \cite{Mezghani_2009, Mezghani_2008_G, Usman_2016,Saxena_2016,Saxena_2016_2, Kakkavas_2016, li_2017, amodh_2020}, several nonlinear precoding algorithms based on different criteria has been proposed in flat fading channel {\cite{Jedda_2016,Swindlehurst_2017,Jacobsson_2016,Jacobsson_2016_2,Jedda_2017,Jedda_2017_2,Jedda_2018,Swindlehurst_2018,Landau_2017,Shao_2018,Jacobsson_2017_2,Wang_2018,Jacobsson_2018,Nedelcu_2018,Tsinos_2018,Ang_Li_2018}}. In this context, the uncoded symbol error rate (SER)  is a widely used design criterion  \cite{Jedda_2016,Swindlehurst_2017,Jedda_2017,Jedda_2017_2,Jedda_2018,Swindlehurst_2018,Landau_2017,Shao_2018,Shao_2019,Li_2020} and is claimed to be more appropriate than the mean squared error (MSE) criterion. Apart from the fact that the impact of channel coding is not taken into account, the superiority of the uncoded SER criterion as compared to the MSE has not been justified adequately, nor has a fair comparison in the large system limit been attempted. Additionally, prior work {including our previous contribution \cite{Jedda_2018}} has focused mainly on frequency-flat channels and has not addressed the spectral shaping issues with low resolution DACs and therefore, skepticism still persists, particularly in the industry, on the appropriateness of the concept.  

{
Solving nonlinear precoding problems, whether based on the MSE or the SRR criterion, is mathematically very challenging. There are three major categories of precoding  algorithms that can be found in the literature. First, a dynamic programming methodology based on the branch-and-bound technique has been adopted in some prior work to solve the problem optimally \cite{Landau_2017,Lopes_2020} or nearly optimally in  \cite{Jacobsson_2018, Li_2020}. These methods are generally not scalable due to the exponential worst-case complexity. Second, convex relaxation has been applied in other related work \cite{Jacobsson_2016,Ang_Li_2018,Jedda_2018,Jacobsson_2017_2} to cope with the non-convex DAC and constant envelope constraints and solve the problem using convex optimization methods. Third,  non-convex local optimization algorithms such as gradient-based methods have been proposed in \cite{Swindlehurst_2017,Tsinos_2018,Wang_2018,Nedelcu_2018,Shao_2018,Shao_2019,Bereyhi_2018}. While not providing rigorous theoretical guarantees, the latter local search optimization techniques often perform better than convex relaxation approaches in practice \cite{Shao_2019}. 
}

In this paper, we formulate and solve the problem of nonlinear multi-user precoding with low resolution DACs {and large antenna arrays} based on the MSE as well as the SER within the general framework of message passing algorithms. To better cope with the low resolution effects, we adopt the adaptive constellation extension (ACE) technique, which was first proposed in the context of peak power reduction in OFDM systems \cite{Krongold}. 
We show by simulation as well as analytically that the MSE based precoder exploiting the ACE technique performs nearly as well as the precoding designed based on the SER  criterion in terms of uncoded SER.

We then exploit the more tractable MSE  criterion combined with ACE to jointly precode and spectrally shape the transmit signals over frequency selective channels {under temporal oversampling}. Confining the signal within the desired frequency band is a critical task in view of the low resolution DAC and the potentially poor out-of-band (OOB) radiation performance \cite{Mollen_Heath_2018}. To handle this issue,  we define the spectral shape over-the-air (OTA) at the receivers' locations instead of the conventional definition at the transmitting antenna ports. Such an approach is meaningful assuming that other users using adjacent channels are co-located with the intended receivers. We also consider an approach to reduce the total unwanted radiation consisting of spatial oversampling beyond the critical half-wavelength antenna spacing.
 
The formulated MSE and SER based combinatorial precoding problems are solved sub-optimally as a graphical model, where belief propagation (BP) or message passing algorithms can be used to solve the optimization iteratively with little complexity per iteration. In particular, the Generalized Approximate Message Passing (GAMP) algorithm \cite{rangan} is an efficient second-order approximation of BP with linear mixing. It is a popular technique {for large scale optimization and estimation problems} in areas such as signal/image processing, statistical physics, and machine learning.
The message passing interpretation of the precoding problem allows us to explore and characterize the performance in the large system limit through the (mathematically non-rigorous) state evolution {(SE)} analysis, known as the cavity method in statistical physics. The large system analysis via the state evolution of GAMP  is a powerful approach and appears to be useful in cases where it seems hopeless to obtain rigorous results from random matrix theory.

Large system analysis has been done in \cite{Sedaghat_2018} based on the replica technique, again another non-rigorous but very powerful and well-known tool from statical physical that appears to be successful in cases where it seems hopeless to obtain rigorous results. The replica technique appears to be quite technical and hardly interpretable. By contrast,  the state evolution analysis provided in this work is strongly linked to the GAMP algorithm and provides therefore much better interpretability of the results based on a simple equivalent scalar model of the system. 
The theoretical grounding for the correctness (or incorrectness) of the SE remains, however, an open area of research. Fortunately, the results predicted by the SE are shown to be rigorous for generalized linear mixing problems including our case under Lipschitzness of the involved proximal/denoising operators and IID assumption on the mixing matrix \cite{Bayati_2011,Barbier5451}. 

 Recently, some publications have proposed the use of the GAMP also to deal with the precoding problem \cite{Chen_2016,Lyu_2017,Bereyhi_2018}. None of this work, however, has considered the state evolution analysis or addressed the DAC issues.  Accordingly, the main contributions of our work are as follows.
 \begin{itemize}
\item We solve the problem of multi-user precoding under low resolution phase DACs using the GAMP algorithm based on both the MSE and SER criteria. Contrary to prior work \cite{Jacobsson_2016_2,Jedda_2017,Jedda_2017_2,Swindlehurst_2018,Jacobsson_2017_2,Wang_2018,Jacobsson_2018,Nedelcu_2018,Jedda_2016,Swindlehurst_2017,Landau_2017,Shao_2018,Shao_2019,Li_2020}, these two criteria are treated with the same algorithmic approach, in which we propose a stochastic relaxation to deal with the non-differentiable input proximal function. The approach is also extended to perform beamforming and spectral shaping in frequency selective channels simultaneously. The constructed waveforms are defined over-the-air (OTA) instead of at each individual antenna port, to exploit the vast amount of degrees of freedom available in massive MIMO. The spectral confinement aims at reducing the adjacent channel interference experienced by co-located-users rather than at the base station. Spatial oversampling is also studied and shown to significantly reduce the total unwanted radiations due to the low resolution effects. 
\item We use the approximate message passing algorithm to track the system performance in the limit of a large number of users and antennas at a fixed ratio. We perform the state evolution analysis of the algorithm and characterize the large system performance by simple fixed point equations for both criteria. Additionally, under the assumption that the state evolution is correct, an equivalent general scalar model is obtained.
\item We demonstrate through simulations as well as analytically that the uncoded performance of the SER criterion can be approached using the MSE criterion combined with ACE. This answers the question disputed in the recent literature of which criterion is best.  Due to its simpler mathematical structure, the MSE criterion is selected for formulating and solving the joint precoding and spectral shaping problem. 
\item We evaluate the performance in terms of OOB radiation measured OTA at the users' locations and show that adequate results can be obtained even with just quadrature-phase DACs. We further demonstrate the benefits of spatial oversampling (such as quarter-wavelength element spacing) for reducing the total radiated OOB power. This addresses the common skepticism regarding the appropriateness of low resolution DACs for generating spectrally confined signals. 
\end{itemize} 

The rest of the paper is organized as follows. In Section II, we describe the system model and formulate both precoding criteria, MSE and SER, for frequency flat channels. In Section III, we present the algorithmic solution based on GAMP. We then perform the state evolution analysis of GAMP in the large system limit in Section IV and provide a scalar channel interpretation. In Section V, we extend our approach to the joint precoding and spectral shaping problem for frequency-selective channels, where we use an OTA definition of the spectrum as observed at the user location. Finally, simulation
results are presented in Section VI. 

\emph{Notation:}
Vectors and matrices are denoted by lower and upper case italic bold letters.  The operators $(\bullet)^\mathrm {T}$, $(\bullet)^\mathrm {H}$, $\textrm{tr}(\bullet)$ and $(\bullet)^*$ stand for transpose, Hermitian (conjugate transpose), trace, and complex conjugate.  The terms $\B{1}_M$ and $\B{I}_M$ represent the all ones vector and the identity matrix of size $M$. { The vector ${\bf e}_i$ is the $i$-th column of the identity matrix}. The vector $\boldsymbol{x}_i$ denotes the $i$-th column of a  matrix $\B{X}$ and $\left[\B{X}\right]_{i,j}$ denotes the ($i$th, $j$th) element, while $x_i$ is the $i$-th element of the vector $\B{x}$.  We represent the Hadamard (element-wise) and the Kronecker product of vectors and matrices by the operators "$\circ$" and "$\otimes$". Additionally, the operator ${\rm Diag}(\cdot)$ constructs a diagonal or a block diagonal matrix from a vector or a sequence of matrices. By contrast,  the operator ${\rm diag}(\cdot)$ extracts the diagonal part of a matrix as a vector. Further, $\mathcal{F}\{\bullet\}$ and $\mathcal{F}^{-1}\{\bullet\}$ are the Fourier and inverse Fourier transform operators used for both the continuous and discrete time domain depending on the context, and $\B{F}_{M}$ represents the normalized  DFT matrix of size $M$ with with swapped
left and right halves and $\B{F}_{M}\B{F}_{M}^{\rm H}=\B{I}_M$. We define the projection operator on a certain set $\mathcal{S}$ as
\begin{equation}
{\rm prox}_{\mathcal{S}}(v)= \argmin_{x \in \mathcal{S}} |x-v|^2.
\end{equation}
Derivatives of functions with multiple complex arguments are expressed in terms of Wirtinger derivatives with respect to the first argument as  
\begin{equation}
f'(v,\cdots)=\frac{1}{2} \left(\frac{\partial f(v,\cdots)}{\partial {\rm Re}(v)} -{\rm j} \frac{\partial f(v,\cdots)}{\partial {\rm Im}(v)}   \right).
\end{equation}
\section{System model and problem formulation}

In this section, we describe the underlying channel and system model. For simplicity of exposition, we consider first a narrowband channel model and focus on the multi-user precoding problem. We extend the approach to the wideband case in Section~\ref{sec_freq_sel}, and address the spectral shaping aspect as well. \\
\subsection{System model}
Consider a massive MIMO downlink transmission with $N$ base station antennas and $K$ single-antenna users. The received  signals, $y_1,y_2,\cdots,y_K$, at the users terminals are expressed as
\begin{equation}
 \B{y}=\B{H}\B{x}+\B{{n}},
\end{equation}
where $\B{x}$ is the precoded vector, $\B{H} \in \mathbb{C}^{K\times N}$ represents the channel matrix, and $\B{{n}} \sim \mathcal{CN}(\B{0}, \sigma_{n}^2 {\bf I} )$ is the noise vector. The channel  matrix is assumed to be perfectly known through this paper. The per-antenna RF chain is equipped with  low resolution phase-only DACs, implemented for instance by a multi-phase phase locked loop (PLL).  
Consequently, the values taken by the transmitted vector $\B{x}$ are constrained to the following finite set
 \begin{equation}
    x_n  \in \mathcal{X}=\left\{{\rm e}^{{\rm j} \frac{2\pi}{2^b} (\ell +\frac{1}{2})}:  \ell=0,\cdots,2^{b}-1  \right\}.
    \label{set_X}
 \end{equation}
Note that the power per antenna is normalized to one without loss of generality.
\begin{figure}[]
\centering
\psfrag{Eb/N0}[c][c]{$E_b/N_0$}
\centerline{\includegraphics[width=3.0in]{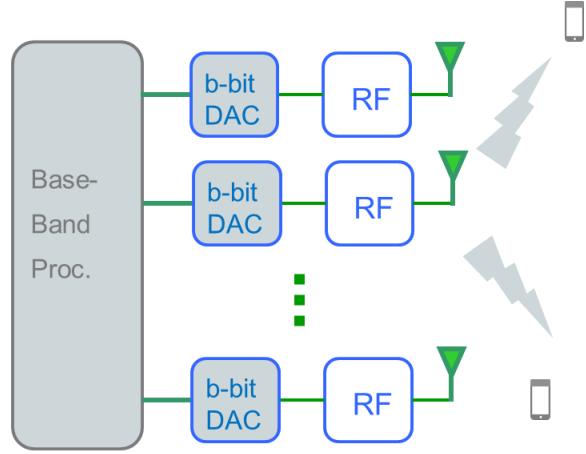}\vspace{-0.0cm}}
\caption{System model for a downlink scenario with $N$ antenna at the base station and $K$ single-antenna users. A per-antenna low resolution DAC is deployed at the transmitter. }
\label{system_model} 
\end{figure}
\subsection{MSE formulation with ACE}

\begin{figure}[htb]
\centering
\psfrag{Eb/N0}[c][c]{$E_b/N_0$}
\centerline{\includegraphics[width=3.4in]{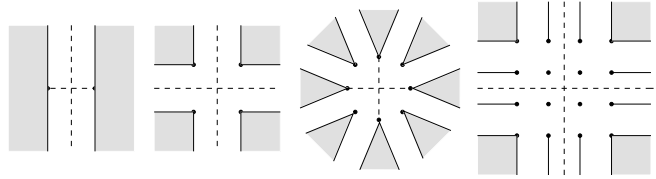}\vspace{-0.0cm}}
\caption{Extended constellations as seen at the receivers without noise, from left to right: BPSK, QPSK, 8PSK, 16QAM.}
\label{extended_cons}
\end{figure}

Adaptive constellation extension (ACE) \cite{Krongold} for digital modulation consists of extending outer constellation points to a convex set while maintaining the minimum distance, i.e., without affecting the symbol error performance. The idea is illustrated in Fig.~\ref{extended_cons} for several type of modulation alphabets. The constellation extension provides more flexibility for the  symbol-by-symbol precoding. This technique is also known as constructive interference (CI) symbol-by-symbol precoding in the more recent literature {\cite{Masouros_2015,Jedda_2017,Jedda_2017_2,Ang_Li_2020,Ang_Li_2021}}.  Mathematically,  the extended set $\mathcal{S}(d)$ of a constellation point  is defined as the convex set
\begin{equation}
  \mathcal{S}(d)=\left\{s:{\rm Re} \{ (s-d)(d-d')^*\} \geq 0, \textrm{for all constellation points } d'\right\}. 
  \label{ex_set}
\end{equation}
 For the case of inner constellation points, we have then simply $\mathcal{S}(d)=d$. Since the distance of the extended set is enhanced to all other points, ACE does not increase the detection error probability.

Assuming a common gain factor $\beta$ at the receiver side{\footnote{ The receiver gain factor $\beta$ is optimized at the base station along with $\B{x}$.  To avoid the impractical overhead for communicating this factor to the users at each transmission, a generalization to block processing is required as discussed later in Section~\ref{sec_freq_sel}. In such a case, once the optimization is solved and the base station has selected the transmit vector, each user can estimate blindly its individual gain factor without overhead as shown in \cite{Jedda_2018}.}}, we can formulate the minimization of the sum MSE for finding the symbol-by-symbol precoding vector $\B{x}$ and the ACE vector $\B{s}$  as the following 
\begin{equation}
\begin{aligned}
\argmin\limits_{\B{s},\B{x}, \beta} \left\| \beta \B{H} \B{x}  -    \B{s} \right\|_2^2 + K \beta^2 \sigma_{n}^2   \quad {\rm s.t.} \quad \B{s} \in \mathcal{S}(\B{d}),~ \B{x} \in \mathcal{X}^N,
\end{aligned} 
\label{MSE_form}
\end{equation}
for given data vector $\B{d}$ to be communicated.

\subsection{SER problem formulation}
An alternative precoding approach is to directly minimize the symbol error probability/rate {under hard-decision detection}. It is mathematically less tractable but can be solved using the same GAMP algorithm described in the next section.  
For the sake of simplicity in the notation, we will focus on the QPSK case with $d_k \in \{\pm 1 \pm {\rm j}\}$; higher {PSK} modulation formats can be also treated with the same concept. 
We can express the conditional probability for detecting the desired data vector $\B{d}$ as the product of the conditional probabilities on each receiver dimension
\begin{equation}
\begin{aligned}
P(\boldsymbol{d}|\boldsymbol{x})&= \prod_{\mathcal{C}\in\{{\rm Re},{\rm Im} \}} \prod_{k=1}^{K}P( \mathcal{C}\{d_k\} |\boldsymbol{x})
=\prod_{\mathcal{C}\in\{{\rm Re},{\rm Im} \}} \prod_{k=1}^{K}\Phi\left(\frac{ \mathcal{C}\{d_k\}  \mathcal{C}\{[\boldsymbol{H}\boldsymbol{x}]_{k}\}}{\sqrt{\sigma_{n}^2/2}}\right),
\end{aligned}
\end{equation}
with $\Phi(x)=\frac{1}{\sqrt{2\pi}}\int_{-\infty}^{x}e^{-\frac{t^2}{2}}d{\rm t}$ being the cumulative normal distribution function.
The corresponding log-likelihood problem aiming  at  minimizing the overall detection error  is given by 
\begin{equation}
\begin{aligned}
\min\limits_{\B{x} \in \mathcal{X}^{N}}  \sum_{k=1}^{K}  \sum_{\mathcal{C}\in\{{\rm Re},{\rm Im} \}}  -\log \Phi\left(\frac{ \mathcal{C}\{d_k\}  \mathcal{C}\{[\boldsymbol{H}\boldsymbol{x}]_{k}\}}{\sqrt{\sigma_{n}^2/2}}\right)  .
\label{SER_form}
\end{aligned}
\end{equation}
It can be noticed that the SER criterion inherently exhibits the extended constellation feature, as any point received in the extended region leads to correct detection. Since QPSK is considered in this case,  the gain factor $\beta$ at the receivers is irrelevant for the optimization. 
\subsection{Dealing with the unwanted spatial harmonics}
Due to nonlinear hardware impairments with large arrays, the generated downlink signal tends to have also dominant spatial emissions besides the desired users channels \cite{Sandrin_1973}, particularly when $K \leq \sqrt[3]{N}$ \cite{mollen_2018}. In other words, the unwanted (out-of-beam) radiated signals $\B{H}_{\perp}\B{x}$, with $\B{H}_{\perp} \in \mathbb{C}^{(N-K)\times N}$ spanning the right null-space of $\B{H}$,  can have still a covariance matrix with dominant eigenvalues, i.e., parasitic spatially focused spurious signals/harmonics \cite{mollen_2018}  (similar to strong sidelobes due to the abrupt signal transitions). This might cause strong interference to other users in the system. One possibility to deal with this issue is to add an anti-sparsity regularization in the cost function based on the spectral norm $\|\B{H}_{\perp} [\B{x}_1 \cdots \B{x}_M] \|_2^2$, with $[\B{x}_1 \cdots \B{x}_M]$ being a matrix containing a certain transmitted sequence of length $M$.  Such regularization term, however, would lead to a significant algorithm complication.
Fortunately, as shown in \cite{mollen_2018}, for a sufficiently large number of users, $K  \geq \sqrt[3]{N}$, the unwanted emissions become inherently isotropic due to the third-order intermodulation products exceeding the number of antennas. In such a case, anti-sparsity regularization is not required.  

Regulatory restrictions do not only limit the unwanted (such as out-of-band) radiations in terms of Effective Isotropic Radiated Power (EIRP), but also in terms of out-of-band (OOB) total radiated power (TRP) \cite{report_ECC}. In such a case, isotropically spreading the unwanted radiation over the space, while being helpful in terms of OOB EIRP, does not reduce the OOB TRP caused by the coarse DAC resolution regardless of the precoding technique,  if no further measures are taken. Possible remedies to this issue could be temporal or spatial oversampling beyond the critical half-wavelength spacing. In fact, if dense spatial sampling is used with antenna spacing below half-wavelength, then the radiated power is not related to the transmit vector norm $\|\B{x}\|$, but instead to a quadratic form involving the conductive/resistive part $\B{B} \in \mathbb{R}^{N\times N}$ of the admittance/impedance matrix of the antenna array \cite{Ivrlac:2010_2}
\begin{equation}
P_{R}=\mathbb{E}[\B{x}^{\rm H} \B{B} \B{x}]={\rm tr} \left(\B{B} \mathbb{E}[\B{x}\B{x}^{\rm H}] \right),
\label{radiated_power}
\end{equation}
where $\B{x}$ is assumed to be the excitation voltage/current vector. In case of spatial oversampling, $\B{B}$ is inherently non-diagonal due to the antenna coupling among densely spaced antennas. Matrix $\B{B}$ is obtained by integrating the outer product of the array angular response vector over the surface of a far field sphere around the antenna array. For a $\sqrt{N} \times \sqrt{N}$ uniform planar array, it is obtained as \cite{williams2019communication}  
 \begin{equation}
\begin{aligned}
&[\B{B}]_{k+\sqrt{N}(\ell-1), m+\sqrt{N}(n-1) }  = 
  \begin{cases}
    \frac{\pi a^2}{\lambda^2}, & \text{for } k=m \text{ and } \ell=n \\
    \frac{a}{\lambda} \frac{ J_1(\frac{2\pi a}{\lambda}\sqrt{(k-m)^2+(\ell-n)^2})}{\sqrt{(k-m)^2+(\ell-n)^2}}, & \text{otherwise, }  
  \end{cases}  
\end{aligned}
\end{equation}
where $J_1(\bullet)$ is the  Bessel function of first kind and first order, $\lambda$ is the wavelength, and $a$ is the antenna spacing.
 The impedance kernel has a step-shaped eigenvalue spectrum \cite{williams2019communication} and acts as a spatial low-pass filter that automatically attenuates certain unwanted emissions, particularly the high spatial harmonics. In other words, a large portion of the unwanted signal power remains stored in the reactive field and does not propagate. This substantially reduces the low resolution artifacts as demonstrated in the simulation results. 
\section{The min-sum-GAMP algorithm}
The min-sum-GAMP algorithm is intended to solve optimization problems involving a sum of numerous similar terms and a linear mixing of the vector to be optimized, as in (\ref{SER_form}) as well as in (\ref{MSE_form}). Due to its performance and generality, we adopt it for sub-optimally solving both non-convex problems.     
The algorithm was derived in slightly different forms in previous work \cite{rangan,Mezghani_WSA_2012,parker1} and is therefore omitted in this paper.  For more historical insights, we also encourage interested readers to consider the much earlier original work in statistical physics \cite{Mezard_1989}, where the GAMP iteration is referred to as the  Thouless-Anderson-Palmer (TAP) equations. {Although known since 1980s, the application of these TAP  equations as an algorithm was only appreciated in the last ten years}. 

\subsection{The GAMP algorithm applied to the precoding problem}
The pseudo-code of GAMP is provide in Algorithm~\ref{GAMP_alg}. The recursive approach at each iteration $\ell$ breaks apart the entire optimization problem into smaller scalar optimizations described by the input function  
\begin{equation} 
   f_\ell(v,\xi^\ell)=  \argmin\limits_{x \in \mathcal{X}}   \frac{1}{\xi^\ell} | \xi^\ell x-v | ^2 = {\rm prox}_{\mathcal{X}} \left( \frac{v}{\xi^\ell} \right)
   \label{f_ell}
\end{equation} 
that incorporates the constraint on $\B{x}$, and  the output function\footnote{We use $-u$ as argument for the function for reasons related to the large system analysis later.} 
\begin{equation} 
   g_\ell(-u,d,\theta^\ell)= \frac{1}{\theta^\ell} \argmin \limits_w  \left[ \Gamma(w,d) +\frac{1}{\theta^\ell} | w-u|^2 \right] - \frac{u}{\theta^\ell}
 \label{out_func}
\end{equation} 
which is related to the cost function, i.e., MSE or SER in our case. More precisely, we have
\begin{equation} 
 \Gamma(w,d)  \stackrel{\rm SER}{=}   \sum_{\mathcal{C}\in\{{\rm Re},{\rm Im} \}}  -\log \Phi\left(\frac{ \mathcal{C}\{d\}  \mathcal{C}\{ w\}}{\sqrt{\sigma_{n}^2/2}}\right), 
 \label{SER_gamma}
\end{equation}  
for the MSE formulation (\ref{SER_form}) with ACE, whereas
\begin{equation} 
 \Gamma(w,d)  \stackrel{\rm MSE}{=}  \min\limits_{s \in \mathcal{S}(d)}  |  w-s | ^2, 
\end{equation} 
for the SER formulation (\ref{MSE_form}) involving finding an optimal ACE point. In the later case, we get the closed form expression 
\begin{equation} 
g_\ell(-u,d,\theta^\ell) \stackrel{\rm MSE}{=}  \frac{ {\rm prox}_{\mathcal{S}(d)}(u) -u}{1+\theta^\ell}.
\label{g_ell}
\end{equation} 
For the SER case, one can use the following approximation based on the Taylor expression of second order of $\Gamma(w,d)$ in (\ref{SER_gamma}) around $w=u$
\begin{equation} 
 g_\ell(-u,d,\theta^\ell)  \stackrel{\rm SER}{\approx} -\frac{\Gamma'(u,d)}{\Gamma''(u,d)+\theta^\ell}.
\end{equation}
The scalar functions $f_\ell$ and $g_\ell$, also called input and output steps \cite{rangan},  are applied element-wise to vectors in the GAMP algorithm.
The messages exchanged between the input and output steps consist, however, not only of the results of the individual scalar optimizations but also the curvature around these optima, which is crucial for faster convergence. The curvature message vectors $\boldsymbol{\xi}^{\ell}$ and $\boldsymbol{\theta}^{\ell}$  are obtained by means of the derivatives $f'_\ell$ and $g'_\ell$  with respect to the first argument. 
It should be noted that the GAMP does not provide guarantees for optimality or convergence.  A damping strategy (also known as averaged operator) is used to enforce the convergence in our case as proposed in \cite{Rangan_GAMP_2014}. 
The recently developed Vector Approximate Message Passing algorithm (VAMP) \cite{rangan_vamp} could be also considered as an alternative to provide better convergence behavior for general (non-IID) channel matrices, but requires the singular value decomposition of the channel matrix and thus high complexity for large system size. Another issue, which we address in the next subsection,  is the fact that the derivative of $f_\ell$ in (\ref{f_ell}) vanishes almost everywhere since the set $\mathcal{X}$ is discrete. Finally, the algorithm can be extended to incorporate an update for the receivers' scalar $\beta$ within its iteration, which will be discussed in Section~V under the more general setting of frequency selective channel.   

\begin{algorithm}[t]
\caption{Damped min-sum-GAMP algorithm}
\label{GAMP_alg}
\begin{algorithmic}[1]
\STATE \textbf{Input:}  data vector $\B{d}$, $\B{A}=\beta\B{H}$ ($\beta=1$ for the SER criterion with QPSK. {The update rule of $\beta$ for the MSE criterion is considered later in Section~\ref{sec_freq_sel} under block processing.}) 
\STATE \textbf{Initialize:} $\boldsymbol{z}^0=\boldsymbol{0}$,
$\boldsymbol{x}^0=\boldsymbol{0}$, $\boldsymbol{\theta}^0={\rm
const} \neq 0$
\\     
$\mu > 0 $,  $\ell \leftarrow 0$      
\REPEAT
\STATE $\ell \! \leftarrow \!  \ell+1$
\STATE Output Step:\\
{$\boldsymbol{z}^{\ell} \! \leftarrow \! g_\ell( - \boldsymbol{A}
{\boldsymbol{x}}^{\ell-1}  + \boldsymbol{\theta}^{\ell-1} \circ
\boldsymbol{z}^{\ell-1},\boldsymbol{d},\boldsymbol{\theta}^{\ell-1})$\\
$\boldsymbol{\xi}^{\ell} \! \leftarrow \! (\boldsymbol{A} \circ \boldsymbol{A}^*)^{\rm T}
g'_\ell ( -\boldsymbol{A} {\boldsymbol{x}}^{\ell-1}  +
\boldsymbol{\theta}^{\ell-1} \circ \boldsymbol{z}^{\ell-1}, \boldsymbol{d},\boldsymbol{\theta}^{\ell-1}) $} \\
\STATE Input Step with damping: \\
{$\boldsymbol{x}^{\ell}  \! \leftarrow \! (1-\mu)\boldsymbol{x}^{\ell-1}+ \mu f_\ell(\boldsymbol{A}^{\rm H}
\boldsymbol{z}^\ell  + \boldsymbol{\xi}^\ell \circ
{\boldsymbol{x}}^{\ell-1},  \boldsymbol{\xi}^\ell)$ \\
$\boldsymbol{\theta}^\ell \! \leftarrow \! \mu  \cdot  (\boldsymbol{A} \circ \boldsymbol{A}^*) 
f'_\ell(\boldsymbol{A}^{\rm H} \boldsymbol{z}^\ell  +
\boldsymbol{\xi}^\ell \circ {\boldsymbol{x}}^{\ell-1},  \boldsymbol{\xi}^\ell )$}
\UNTIL{the cost does not significantly decrease or a maximum iteration count has
 been reached}
\STATE Final projection:\\
   $\B{x}_{\rm sol}={\rm prox}_{\mathcal{X}}(\B{x}^{\ell})$
\end{algorithmic}
\end{algorithm}

\subsection{A stochastic relaxation technique for the min-sum-GAMP}
\begin{figure}[htb]
\centering
\psfrag{v}[c][c]{$v$}
\psfrag{f(v)}[c][c]{$f(v)$}
\centerline{\includegraphics[width=2.4in]{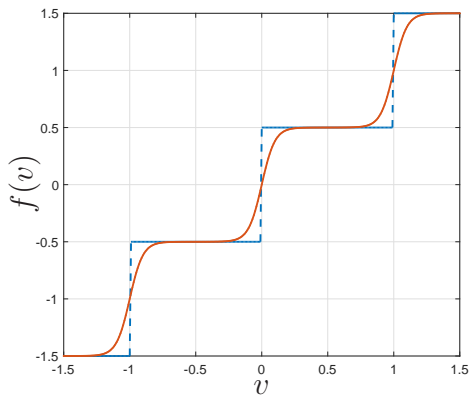}\vspace{-0.0cm}}
\caption{Stochastic relaxation of a staircase function using the \emph{Softmax} function (\ref{softmax}).}
\label{stochastic_relaxation}
\end{figure}

To cope with the non-smoothness of the function $f_\ell(v,\xi)$ in (\ref{f_ell}), we approximate the minimization as averaging from a Gibbs distribution. This can be expressed as (also know as the \emph{Softmax} function)
\begin{equation}
f_\ell(v,\xi^\ell)\approx  \frac{\sum\limits_{x \in \mathcal{X}} x
{\rm e}^{-\frac{\gamma^\ell}{\xi^\ell} |\xi^\ell x-v|^2} }{\sum\limits_{x \in \mathcal{X}}
{\rm e}^{-\frac{\gamma^\ell}{\xi^\ell} |\xi^\ell x-v|^2} },
\label{softmax}
\end{equation}
where $\gamma^\ell$ is an artificial inverse temperature parameter and is increased gradually as the iteration evolves, corresponding to a simulated annealing procedure. When $\gamma^\ell$ approaches infinity, the distribution concentrates into the exact solution in (\ref{f_ell}). The smoothing effect via this technique is also known as stochastic relaxation and provides better and more robust performance of the original min-sum-GAMP algorithm. An illustrative example is given in Fig.~\ref{stochastic_relaxation}.

Another alternative to cope with the non-smoothness issue it to use the approximation
\begin{equation} 
f'_\ell(v,\xi^\ell) \approx  \frac{1}{2|v|},
\end{equation}
which becomes the exact Wirtinger derivative of $f_\ell(v,\xi)=v/|v|$ for the infinite resolution case with only a constant envelope constraint. 
\section{Large system analysis}
\label{Large_system}
A large system analysis aims at characterizing the performance of the considered system when the number of antennas and the number of users grow to infinity at a fixed ratio of $K/N$. To this end, we will assume IID channel coefficients with a variance of $1/N$. Such analysis has been done recently in \cite{Sedaghat_2018} based on the replica technique, a popular yet nonintuitive (from our perspective) and mathematically non-rigorous method from statistical physics. In the following, we provide an alternative approach based on the state evolution (also known as the density evolution
for the case of the channel coding) framework of the GAMP algorithm. This framework leads to an equivalent scalar model governed by scalar equations characterizing the large system performance similarly to the replica analysis but in a more intuitive way. 

The derived performance equations are still mathematically non-rigorous due to some empirical assumptions, know as the replica symmetry hypothesis in statistical physics \cite{Mezard_1989},  that are discussed later on. The validity of these assumptions is a challenging open problem in general. Nevertheless, rigorous proofs of the state evolution analysis have been developed in \cite{Bayati_2011,Barbier5451} for the case of random IID matrix $\B{H}$ and \emph{Lipschitz} functions $f$ and $g$  (Note that $f$ is not \emph{Lipschitz} in our case due to the quantization). This is consistent with the claim made in \cite{Sedaghat_2018} that the replica technique, and thus the state evolution method as well, provide correct results in the case of infinite resolution DACs with constant envelope constraint. In the finite resolution case, however, the predicted results are claimed to be incorrect. Nevertheless, a comparison to the results obtained by simulation shows that these theoretical predictions are still useful in the lower SNR ranges.

\par For the state evolution of Algorithm~\ref{GAMP_alg}, we make the assumption of $\mu=1$ (no damping) and that the vectors  $\B{u}^{\ell}=\beta\boldsymbol{H}{\boldsymbol{x}}^{\ell-1}-\boldsymbol{\theta}^{\ell-1} \circ\boldsymbol{z}^{\ell-1}$ and $\B{v}^{\ell}=\beta\boldsymbol{H}^{\rm H}\boldsymbol{z}^\ell  + \boldsymbol{\xi}^\ell \circ {\boldsymbol{x}}^{\ell-1}$ used for calculating the output and the input step respectively, converge to a zero mean IID Gaussian vectors in the large system limit\footnote{The scaling factor $\beta$ is only required for the MSE criterion, and is simply equal 1 for the SER optimization with QPSK.}. This holds true provided the conjecture that incoming messages for each step are independent, which is equivalent to the replica symmetry assumption in statistical physics \cite{Mezard_1989,Sedaghat_2018}. Under this conjecture, the entries of $\B{u}^{\ell}$ and $\B{v}^{\ell}$ converges in distribution with $\ell,~N,~K \longrightarrow \infty$ at fixed $K/N$, to the scalar distributions
\begin{align}
  u &\sim \mathcal{CN}(0, \beta^2 q) \label{u_large}\\
 v &\sim \mathcal{CN}(0, \beta^2 \nu), 
\end{align}
 where
\begin{equation}
\begin{aligned}
    q &= {\rm E}[|f(v,\xi)|^2] \stackrel{\rm Phase~Quant.}{=} 1  \textrm{ and }
 \nu &= \frac{K}{N} {\rm E}[|g(-u,d,\theta)|^2] .
\end{aligned}
\label{state_eq1}
\end{equation}
Note that $q$ is related to the per-antenna power.

Additionally, the parameters $\B{\theta}^{\ell}$ and $\B{\xi}^{\ell}$ in Algorithm~\ref{GAMP_alg} are sums of terms of order $1/N$, thus
they admit asymptotically zero variance. In other words,
they become deterministic in the large-system limit at given
iteration provided that $\B{H}$ has IID entries. We have therefore the asymptotic constant values for the elements of vectors  $\B{\theta}^{\ell}$ and $\B{\xi}^{\ell}$
\begin{align} 
 \theta &= \beta^2 {\rm E}_v[f'(v,\xi)],  \\
 \xi &= \frac{K}{N}  \beta^2 {\rm E}_u[g'(-u,d,\theta)]. 
\end{align}
For the case of phase quantization DAC, where $f(v,\xi)$ is projecting on the set $\mathcal{X}$ defined in (\ref{set_X}), we get using Stein's Lemma {\cite{Stein_1981}}
\begin{align} 
 \theta &= \beta^2 {\rm E}[f'(v,\xi)]= \beta^2 \frac{{\rm E}[f(v,\xi)v^*]}{{\rm E}[|v|^2]} = \frac{2^b \sin(\pi/2^b)}{2\sqrt{\pi \nu/(\beta^2q)}}. \label{state_eq2} 
\end{align}

\begin{figure}[htb]
\centering
\psfrag{xd}[c][c]{$\hat{s}$}
\psfrag{w}[c][c]{$w$}
\psfrag{beta}[c][c]{$\beta$}
\psfrag{x}[c][c]{$d$}
\psfrag{theta}[c][c]{$\theta$}
\psfrag{f(z)}[c][c]{$g(\!-u,\!d,\!\theta)$}
\psfrag{noise1}[c][c]{\tiny $ \!\!\!\!\ u \! \sim \! \mathcal{CN}(0, \beta^2 q )$}
\psfrag{noise2}[c][c]{\tiny $ ~~~~~~~~~~~~~~~  \beta {n} \! \sim \! \mathcal{CN}(0, \beta^2 \sigma_{n}^2 )$}
\centerline{\includegraphics[width=3.4in]{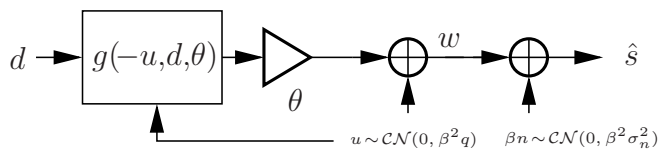}\vspace{-0.0cm}}
\caption{Equivalent scalar model, ($\beta=1$ for the SER criterion).}
\label{scal_equiv}
\end{figure}

The conjectured asymptotic behavior of the message algorithm implies that the performance can be fully described by a one-dimensional scalar model, where the underlying effective channel for each user can be deduced from the output function (\ref{out_func}) as
\begin{equation}
 \hat{s} = w+ \beta {n}= \theta g(-u,d,\theta) + u + \beta {n}.
\end{equation} 
 The equivalent scalar model is illustrated in Fig.~\ref{scal_equiv}. The random variable $u$ can be understood as the resulting effective multi-user interference that is known at the transmitter, which can be taken into account when designing the precoding strategy. In the following, we analyze the performance of the different precoding techniques, namely, the MSE minimization with and without ACE, as well as the SER minimization.

\subsection{MSE criterion without ACE}
The MSE criterion without ACE has been considered already in \cite{Sedaghat_2018} and the state evolution equations (\ref{state_eq1}) and (\ref{state_eq2}) are exactly the same as derived in \cite{Sedaghat_2018} using the different replica method. As shown in the following,  however, the interpretation as a scalar channel model in Fig.~\ref{scal_equiv} provides, a more accurate and insightful characterization of the channel capacity than in \cite{Sedaghat_2018}, where only a capacity lower bound was derived. 
\par For the MSE criterion without ACE, i.e., $\mathcal{S}(d)=d,~\forall d$, the output function $g(-u,d,\theta)$ in (\ref{g_ell}) becomes
\begin{equation}
  g(-u,d,\theta)=\frac{d-u}{1+\theta},
\end{equation}
with (using  (\ref{u_large}))
\begin{equation}
\nu=\frac{K}{N}{\rm E}[|g(-u,d,\theta)|^2]=\frac{K}{N}\frac{\sigma_d^2+\beta^2q}{(1+\theta)^2}.
\label{nu_MSE_no}
\end{equation}
In addition, we obtain from  Fig.~\ref{scal_equiv} the effective SINR for given ${\rm SNR}=q/\sigma_{n}^2$
\begin{equation}
 {\rm SINR}=\frac{\theta^2 \sigma_d^2/(\beta^2 q)}{1+\frac{(1+\theta)^2}{\rm SNR}}.
\end{equation} 
Further, we express $\sigma_d^2/(\beta^2 q)$ as function of $\nu$ using  (\ref{nu_MSE_no}), and then $\nu/(\beta^2 q)$ as function of $\theta$ using (\ref{state_eq2}) with phase quantization, to get the SINR formula as function of only parameter $\theta$:

\begin{equation}
  {\rm SINR}= \frac{\frac{N}{4\pi K}2^{2b}
\sin^2(\pi/2^b)(1+\theta)^2 -\theta^2}{1+\frac{(1+\theta)^2}{\rm SNR}}.
 \label{SINR}
\end{equation}
We can now maximize the SINR with respect to $\theta$ to maximize the performance, which corresponds also to finding the optimal receiver scaling factor $\beta$ due to the relations in (\ref{nu_MSE_no}) and (\ref{state_eq2}). This can be done by finding the positive root of the derivative of (\ref{SINR}) with respect to $\theta$. The maximization leads to the following result, where surprisingly the  ${\rm SINR}(\theta)$ is maximal at its fixed point  
\begin{equation}
\begin{aligned}
&{\rm SINR}_{\rm opt}= \theta_{\rm opt}  
=\frac{a-1}{2}{\rm SNR}-\frac{1}{2}+\sqrt{a{\rm
SNR}+\left(\frac{a-1}{2}{\rm SNR}-\frac{1}{2} \right)^2}, 
\end{aligned}
\label{SINR_opt}
\end{equation} 
with $a=\frac{2^{2b}N}{4\pi K} \sin^2(\pi/2^b)$. Finally, the achievable rate is given by $C=\log_2(1+\theta_{\rm opt})$, which is slightly different than the lower bound obtained in \cite{Sedaghat_2018}. {The uncoded SER is computed according to the modulation type and an AWGN channel specified by this SINR}.
\subsection{MSE criterion with ACE}
The output function $g(-u,d,\theta)$ formulated based on the MSE criterion with ACE is given in (\ref{g_ell}) and it depends on the extended set (\ref{ex_set}) of each constellation point.  As a particular example, we derive the state evolution equations for the case of $L^2$-QAM constellation. First, we derive the parameter $\nu$ defined in (\ref{state_eq1})
\begin{equation}
\begin{aligned}
 \nu &= \frac{K}{N} \frac{1}{L^2} \sum_d \int\limits_\mathbb{C} \min\limits_{s\in \mathcal{S}(d)} \frac{|s- u|^2}{(1+\theta)^2} \frac{{\rm e}^{-\frac{|u|^2}{\beta^2 q}}}{\sqrt{\pi \beta^2 q}} {\rm d}u \\
    &\stackrel{(a)}{=}  \frac{K}{N}\frac{1}{(1+\theta)^2}  \Bigg((1-\frac{2}{L})(\frac{2}{3}((L-2)^2-1)+\beta^2q)  
     + \frac{4}{L} \int\limits_{-\infty}^{(L-1)}  |L-1- u|^2 \frac{{\rm e}^{-\frac{|u|^2}{\beta^2 q}}}{\sqrt{\pi \beta^2 q}} {\rm d}u \Bigg)  \\
    &= \frac{K}{N} \frac{1}{(1+\theta)^2}  \Bigg((1-\frac{2}{L})(\frac{2}{3}((L-2)^2-1)+\beta^2q) \\
    &~~ + \frac{2}{L} \Bigg(\sqrt{\frac{q}{\pi}} \beta {\rm e}^{-\frac{(L-1)^2}{q}}(L-1) 
      + (2(L-1)^2+q \beta^2) \Phi(\frac{L-1}{\beta\sqrt{q/2}}) \Bigg)   \Bigg) ,
\end{aligned}
\label{nu_ACE}
\end{equation}
where in step $(a)$, we distinguish between inner and outer points. From (\ref{nu_ACE}) and (\ref{state_eq2}), we get fixed point equations in terms of $\nu$ and $\theta$, that can be solved numerically. The resulting MSE per user's channel is then given by
\begin{equation}
{\rm MSE}=\frac{N}{K} \nu + \beta^2 \sigma_{n}^2.
\end{equation}
Additionally, we calculate the symbol error probability for the $L^2$-QAM constellation, while distinguishing the contributions of the inner and outer points 
\begin{equation}
\begin{aligned}
{\rm SER}&= 1-\left(1- 2\frac{L-2}{L}
\Phi\left(-\sqrt{\frac{2\theta^2/(\beta^2 q)}{1+{(1+\theta)^2}/{\rm SNR}}}\right) +  \right.  \\
&  \left.  \frac{2}{L} \int\limits_{-\infty}^{\infty}  \Phi\left(  -\frac{u+\theta \max(1,u-L+2) }{(1+\theta)\beta \sigma_{n}/\sqrt{2}} \right) \frac{{\rm e}^{-\frac{u^2}{\beta^2 q}}}{\sqrt{\pi \beta^2 q}} {\rm d}u  \right)^2.
\end{aligned}
\label{SER_ACE_crit}
\end{equation}
Finally, the receiver scaling factor $\beta$ can be optimized to minimize for instance either the MSE or the SER.  
\subsection{SER criterion for the QPSK constellation}
Concerning the SER based precoding with the QPSK constellation, we set $\beta$ to one and we define based on (\ref{out_func}) and (\ref{SER_gamma}) the following output function for each real dimension with $d_R \in\{\pm 1\}$
\begin{equation}
\begin{aligned}
&g(-u,d_R,\theta)= \frac{1}{\theta} \argmax\limits_w \left[   \log
\Phi\left(\frac{ d_R w}{\sigma_{n}/\sqrt{2}}\right) -\frac{(w-u)^2}{\theta} 
\right] - \frac{u}{\theta}.
\end{aligned}
\end{equation}
Consequently, the state-evolution parameter $\nu$ is obtained as
\begin{equation}
\begin{aligned}
 \nu &=2 \frac{K}{N}  \int\limits_{-\infty}^{\infty} g(-u,1,\theta)^2
\frac{{\rm e}^{-\frac{u^2}{q}}}{\sqrt{\pi q}} {\rm d}u.
\end{aligned}
\label{nu_SER}
\end{equation}
For QPSK, the SER is related to the bit error rate (BER) according to ${\rm SER}=2 \cdot {\rm BER}-{\rm BER}^2$. The calculation of the BER based on the scalar channel model in Fig.~\ref{scal_equiv} yields
\begin{equation}
\begin{aligned}
 {\rm BER} =  \int_{-\infty}^{\infty}  \Phi\left(- \sqrt{2}(\theta
g(-u,1,\theta)+u)/\sigma_{n}\right) \frac{{\rm
e}^{-\frac{u^2}{q}}}{\sqrt{\pi q}} {\rm d}u.
\end{aligned}
\label{BER_SER_crit}
\end{equation}
Again, the parameters $\theta$ and $\nu$ of the scalar model are obtained by numerically solving the fixed point equations in (\ref{state_eq2}) and (\ref{nu_SER}).
\section{Joint precoding and spectral shaping in frequency selective channels}
\label{sec_freq_sel} 

 {Temporal processing that is needed for pre-equalizing frequency-selective channels as well as for spectral shaping leads to precoding problems of a much larger dimension than the previous space-only case, calling for more efficient precoding methods}.
We generalize in this section the {MSE} precoding design to a block fading frequency selective channel with $K$ single antenna users  and $N$  {transmitting} antennas in the downlink. During a coherence time of $M$ symbols, the base station processes for instance a block of $M/2$ data symbols using single carrier pulse shaping or OFDM  processing, constructs (with an oversampling factor of 2) the corresponding stacked output vector $\B{x}=[\B{x}^{\rm T}[0],\ldots,\B{x}^{\rm T}[M-1]]^{\rm T} \in \mathcal{X}^{MN}$ for the DACs  and appends a cyclic prefix that is longer than the delay spread. After convolution with the channel, and discarding the cyclic prefix, the {users receive} a block of $M$ sampled signal vectors. Moreover, the matrix sequence $\B{H}[m] =[\B{h}_1[m],\ldots,\B{h}_K[m]]^{\rm T} \in \mathbb{C}^{K \times N}$ comprises the user channels in the DFT domain that are assumed to be  perfectly known. Considering the DFT of the $M-$length discrete-time signals, the sampled and stacked received signal {block} in the DFT domain is
\begin{equation}
    \B{y} =  {\rm Diag} \left(\B{H}\left[-\frac{M}{2}\right]\cdots\B{H}\left[\frac{M}{2}-1\right]\right) \left(\B{F}_{\!M} \otimes \B{I}_{\!N}\right) \B{x}  + \B{{n}} 
    \label{ym}
\end{equation}
where $\B{{n}} \in \mathbb{C}^{NM}$ is the stacked noise vector in the frequency domain having IID elements with unit variance.

The joint  precoding and OTA spectral shaping problem can be formulated in the frequency domain assuming a receiver scaling factor $\beta$ and a desired spectral shaping matrix $\B{G}$ as
\begin{equation}
\begin{aligned}
&\min\limits_{\B{x},\B{s},\beta} \! \left\| \beta {\rm Diag} \left(\B{H}\left[-\frac{M}{2}\right]\cdots\B{H}\left[\frac{M}{2}-1\right]\right) \left(\B{F}_{\!M} \otimes \B{I}_{\!N}\right) \B{x}   \!- \! (\B{G} \otimes \B{I}_{\!K})  \B{s} \right\|_2^2 +   \\
& \quad\quad\quad\quad\quad\quad\quad\quad\quad\quad\quad\quad\quad\quad + KM\sigma^2_{n} \beta^2  \quad {\rm s.t.} \quad \B{x}  \in \mathcal{X}^{MN},~\B{s} \in \mathcal{S}(\B{d}) {\in \mathbb{C}^{\frac{M}{2}K}}.
\end{aligned} 
\label{cost_func}
\end{equation}
The OTA spectral shaping aims at reducing the adjacent-channel interference experienced by potential co-located users. The formulation can be also modified to handle the case of adjacent-channel users at locations other than the served ones.  We focus rather on the case of co-located users as it is natural to separate them in frequency instead of space. This OTA waveform strategy is different than the common approach where the waveform is created and measured at each transmit antenna. We believe that the conventional approach, which has mainly evolved from the single antenna case,  is very restrictive and cannot exploit the potential of massive MIMO for nulling interference at several locations.\footnote{The OTA waveform strategy requires certain coordination between operators serving the same area at the same frequency band. These issues can be resolved simply by sharing the base station hardware and even the entire frequency band.}
The OTA spectral shaping matrix $\B{G}$ shall satisfy the orthogonality condition  $\B{G}^{\rm H}\B{G}={\bf I}_{M/2}$ (i.e., ISI-free or Nyquist pulse).
Each receiver applies the matrix $\B{G}^{\rm H}$ (matched filter) to the received data block to get an estimate of $\hat{\B{s}}$.  Examples of spectral shaping matrices $\B{G}$ for common waveforms at two samples per symbol interval are provided in Table~\ref{uniform}. The diagonal matrix $\B{\Lambda}$ represents the spectral profile in the frequency domain, such as an RRC filter with a certain roll-off. \\

\begin {table}[thp]%
\caption {Examples of spectral shaping matrices $\B{G}$ with 2 samples per symbol}
\label{uniform}\centering %
\begin{tabular}{|c|c|c|}
\hline  %
 OFDM with cyclic prefix (CP) & CP based Single-carrier (CP-SC)  &  Offset-QAM  CP-SC (real valued $\B{d}$)  \\\hline
  $ \left[    
\begin{array}{ll}
	\B{I}_{M/4} & \B{0}_{M/4}  \\
  \B{0}_{M/2,M/4} &  \B{0}_{M/2,M/4}  \\
			\B{0}_{M/4} & \B{I}_{M/4}  \\
\end{array}
\right]$   &  $ \B{\Lambda}  \left[    
\begin{array}{l}
	\B{F}_{M/2}  \\
	\B{F}_{M/2}
\end{array}
\right]$  &   $(\B{\Lambda}  \B{F}_{M})  \cdot  \B{I}_{M/2} \otimes  \left[    
\begin{array}{ll}
1 & 0  \\
0 & {\rm j}
\end{array}
\right]$   \\\hline %
\end {tabular}
\end {table}
To solve the generalized precoding and pulse shaping problem, the output step of the GAMP Algorithm~\ref{GAMP_alg} is changed as follow
\begin{equation}
\begin{aligned}
g_\ell(-\B{u},\B{d},\B{\theta}) = \frac{1}{\theta^\ell} \argmin\limits_{\B{w},\B{s} \in \mathcal{S}(\B{d})} \!\!  \left[ \left\| \B{w}- (\B{G} \otimes \B{I}_{\!K})  \B{s} \right\|_2^2  \!+\! \frac{\left\| \B{w}- \B{u} \right\|_2^2}{\theta^\ell}  \right]\! - \!\frac{\B{u}}{\theta^\ell},
\end{aligned} 
\end{equation}
where we used the approximation $\B{\theta}^\ell \approx {\theta}^\ell \B{1} $.
Using the fact that  $\B{G}^{\rm H}\B{G}={\bf I}$, this minimization can be solved first with respect to $\B{w}$, then {$\B{s}$} after expanding the quadratic from. The final solution is given by
\begin{equation}
\begin{aligned}
& g_\ell(-\B{u},\B{d},\B{\theta})  =   {\rm Diag}(\B{1}+\B{\theta}^\ell)^{-1} \left((\B{G} \otimes \B{I}_K){\rm prox}_{\mathcal{S}(\B{d})} \left((\B{G} \otimes \B{I}_K)^{\rm H}\B{u}\right) - \B{u}\right).
\end{aligned} 
\label{gl_freq_sel}
\end{equation}
The curvature vector $\B{\xi}^\ell$ can be expressed as
\begin{equation}
\begin{aligned}
\B{\xi}^\ell = {\rm diag}( \B{A}^{\rm H} \nabla_{\B{u}} g_\ell(-\B{u},\B{d},\B{\theta})   \B{A}  ) ,
\end{aligned} 
\end{equation}
with $\B{A} = \beta {\rm Diag}(\B{H}[m]) (\B{F}_{\!M} \otimes \B{I}_{\!N})$, and the Jacobian matrix
\begin{equation}
\begin{aligned}
\!\! & \nabla_{\B{u}} g_\ell(-\B{u},\B{d},\B{\theta}) = {\rm Diag}(\B{1}+\B{\theta}^\ell)^{-1} \cdot  \left(\B{I}\!-\!(\B{G} \otimes \B{I}_{\!K}\!){\rm Diag}\!\left({\rm prox}'_{\mathcal{S}(\B{d})} \left((\B{G} \otimes \B{I}_{\!K}\! )^{\!\rm H}\B{u}\right)\right) (\B{G} \otimes \B{I}_{\!K}\!)^{\!\rm H} \right) \!.
\end{aligned} 
\end{equation}
Finally the update for $\beta$ for fixed precoder output {$\B{x}$} is obtained from solving (\ref{cost_func}) with respect to $\beta$ in closed form as
\begin{equation}
\begin{aligned}
\beta = \frac{\B{s}^{\rm H} (\B{G} \otimes \B{I}_K)^{\rm H} {\rm Diag}(\B{H}[m]) (\B{F}_M \otimes \B{I}_N) \B{x}}{ {\|\rm Diag}(\B{H}[m]) (\B{F}_M \otimes \B{I}_N) \B{x} \|_2^2+KM \sigma_n^2},
\end{aligned} 
\label{beta_opt}
\end{equation} 
which can be performed within the recursion of Algorithm~\ref{GAMP_alg}.  The initial value for  $\beta$ can be chosen equal one.
\section{Simulation results}
\label{simulation}
In this section, we provide some numerical results for the generic IID channel model, as well as for a practical wideband
mmWave setting. In our setting, no power control is performed by the base station, as it is not included in current cellular standards for the downlink\footnote{In the downlink, rate control is commonly used instead of power control.}. Therefore, we assume, without loss of generality,  that the users have the same pathloss. In all simulations, the proposed algorithm converges typically within 20 to 30 iterations. The results we report are the averages over 100 channel realizations.
\subsection{SER vs. MSE criterion and analytical vs. Monte Carlo simulation for IID flat fading channels}

\begin{figure}[htb]
\centering
\psfrag{xd}[c][c]{$\hat{s}$}
\centerline{\includegraphics[width=3.5in]{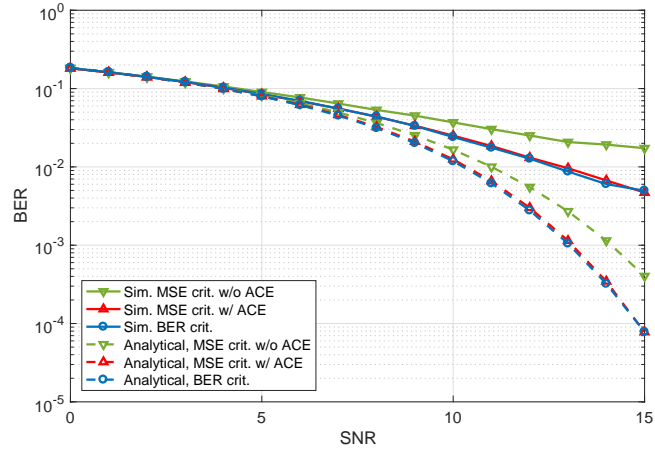}\vspace{-0.0cm}}
\caption{BER performance using the MSE as compared to the SER criterion for $N=40$, $K=20$, QPSK constellation  and 2-bit phase DACs. Note that SER and BER minimization are equivalent for QPSK. The performances of the ACE-based MSE and SER criteria are similar, which is also predicted by the SE analysis.}
\label{sim1}
\end{figure}

We first consider an IID flat fading channel model to be able to verify the empirical state evolution analysis. In Fig.~\ref{sim1}, a performance comparison between the MSE and SER based precoders is shown, where the base station is equipped with $N=40$ antennas and serves $K=20$ users using QPSK constellation and 2-bit phase DACs. It can be seen that the MSE based performance can approach the SER performance when ACE is utilized. The MSE criterion combined with ACE is therefore advantageous over the SER criterion due to its simpler mathematical formulation. The same observation can be made based on the theoretical results from the state evolution analysis {(c.f. (\ref{SINR_opt})
, (\ref{SER_ACE_crit}), and (\ref{BER_SER_crit}))}.  In fact, these analytical results provide satisfactory performance prediction up the medium SNR regime and are also useful for performance comparison.

\begin{figure}[htb]
\centering
\psfrag{xd}[c][c]{$\hat{s}$}
\centerline{\includegraphics[width=3.5in]{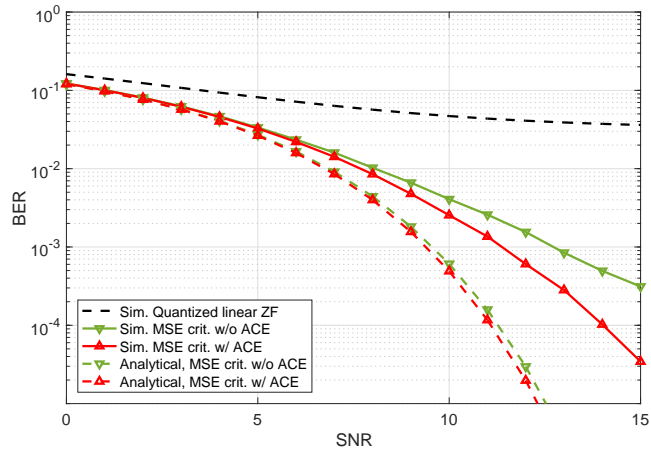}\vspace{-0.0cm}}
\caption{BER performance using the MSE as compared to the SER criterion for $N=40$, $K=13$, QPSK constellation and 2-bit phase DACs.  {The analytical results are based on (\ref{SINR_opt}) and (\ref{SER_ACE_crit}).}}
\label{sim2}
\end{figure}

In Fig.~\ref{sim2}, we decrease the number of users to $K=13$ and we compare the proposed nonlinear MSE based precoder to the quantized linear zero-forcing (ZF) precoder. As expected, a substantial improvement can be achieved and linear methods are still insufficient for this setting with a lower number of users.  

\begin{figure}[htb]
\centering
\psfrag{xd}[c][c]{$\hat{s}$}
\centerline{\includegraphics[width=3.5in]{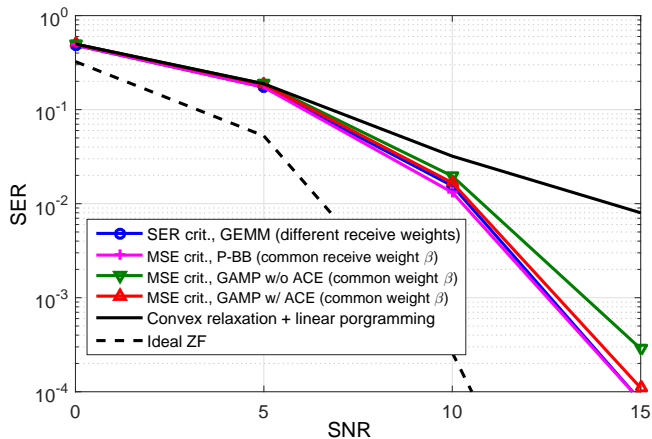}\vspace{-0.0cm}}
\caption{SER performance comparison of the proposed precoding algorithms to other recent methods for $N=128$, $K=16$, 16QAM constellation,  and 2-bit phase DACs. }
\label{sim30}
\end{figure}

The performance with the higher-order modulation, i.e., 16QAM is considered in Fig.~\ref{sim30}, where the number of antennas is set to $N=128$ while the number of users is $K=16$. {Since the receiver gain $\beta$ is now crucial, we optimize it using block-processing similarly to (\ref{beta_opt}) with $M=25$, which allows for blockwise static $\beta$.} We compare our MSE-GAMP based design with and without ACE to recent techniques from the literature namely the gradient extrapolated majorization-minimization (GEMM) method \cite{Shao_2019}, the partial branch-and-bound (P-BB) procedure  \cite{Li_2020}, {and our prior method based on convex relaxation of the DAC constraint and linear programming \cite{Jedda_2018}}. { Combined with ACE, our MSE based technique approaches the performance of the P-BB method that can have exponential worst-case complexity.  It also nearly achieves the same performance as GEMM, which is based on the SER criteria and requires the additional optimization of each user's receive weight individually.} This again confirms our previous observation that the MSE criteria can be as good as the SER criteria when combined with ACE while being mathematically more elegant and flexible.  

\begin{figure}[htb]
\centering
\psfrag{xd}[c][c]{$\hat{s}$}
\centerline{\includegraphics[width=3.5in]{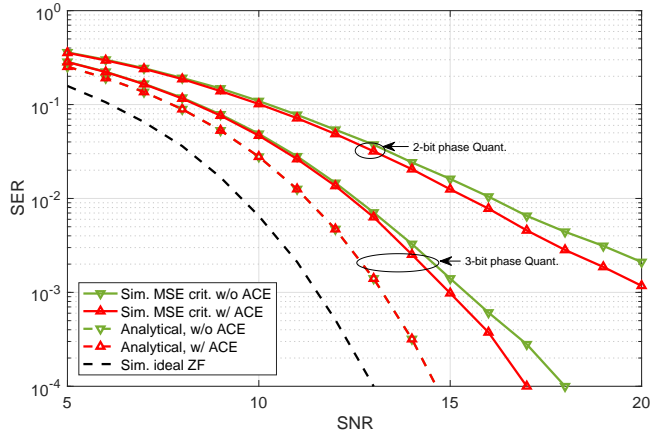}\vspace{-0.0cm}}
\caption{SER performance for $N=100$, $K=20$, 16QAM constellation,  and 2-bit and 3-bit phase DACs. {The analytical results are based on (\ref{SINR_opt}) and (\ref{SER_ACE_crit}).}}
\label{sim3}
\end{figure}

The performance with two different DAC resolutions, $b \in\{2,3\}$, are depicted in {Fig.~\ref{sim3}} for $N=100$ and $K=20$. It can be noticed that 16QAM can be still supported by the quadrature DAC ($b=2$), while increasing the resolution by one bit can provide 2.5dB  SNR improvement at $10^{-2}$ SER. The performance gains from ACE are reduced compared to the QPSK case, as the inner points of the 16QAM do not benefit from ACE (see Fig.~\ref{extended_cons}). 
  
\begin{figure}[htb]
\centering
\psfrag{xd}[c][c]{$\hat{s}$}
\centerline{\includegraphics[width=3.0in]{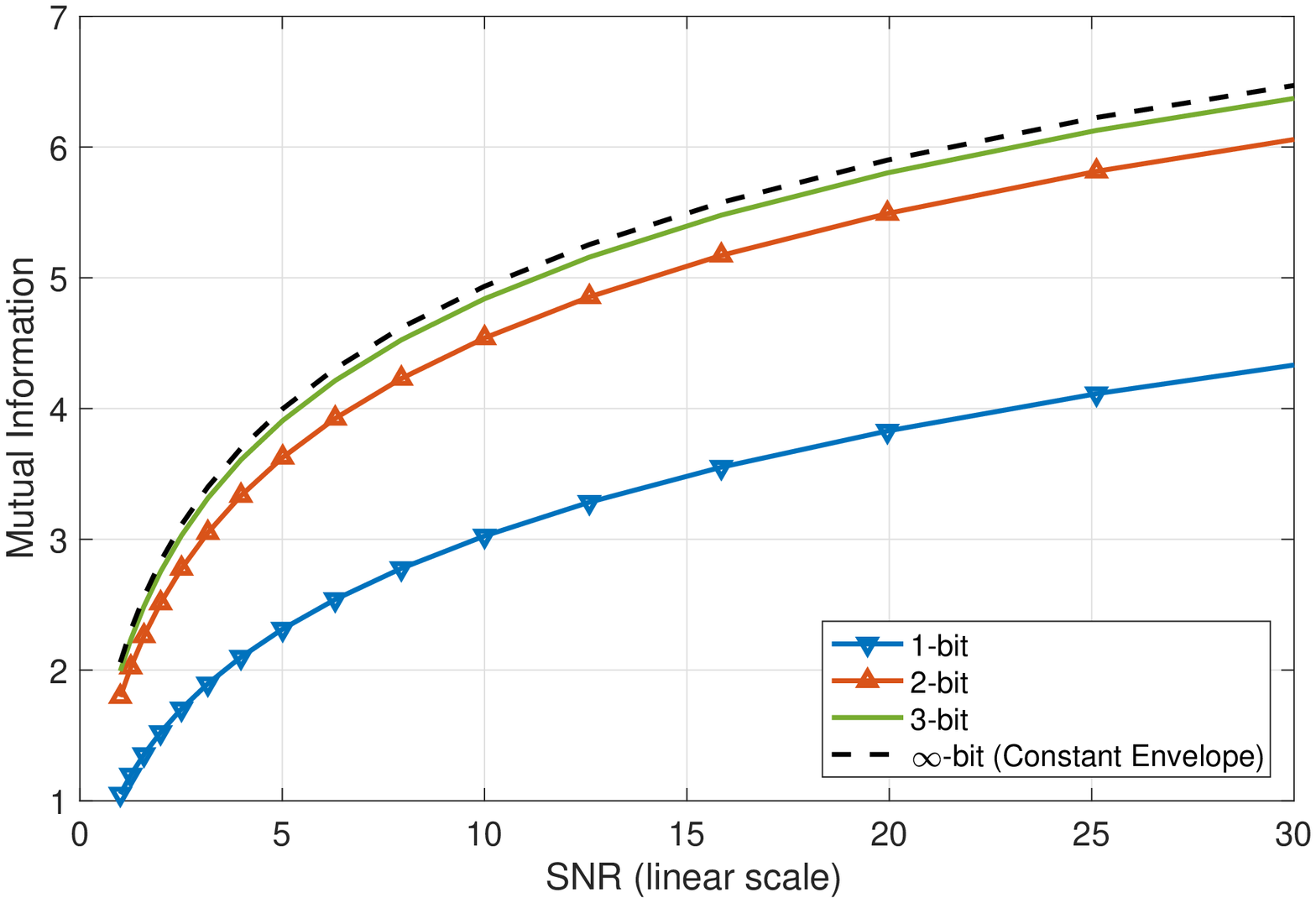} 
\includegraphics[width=3.0in]{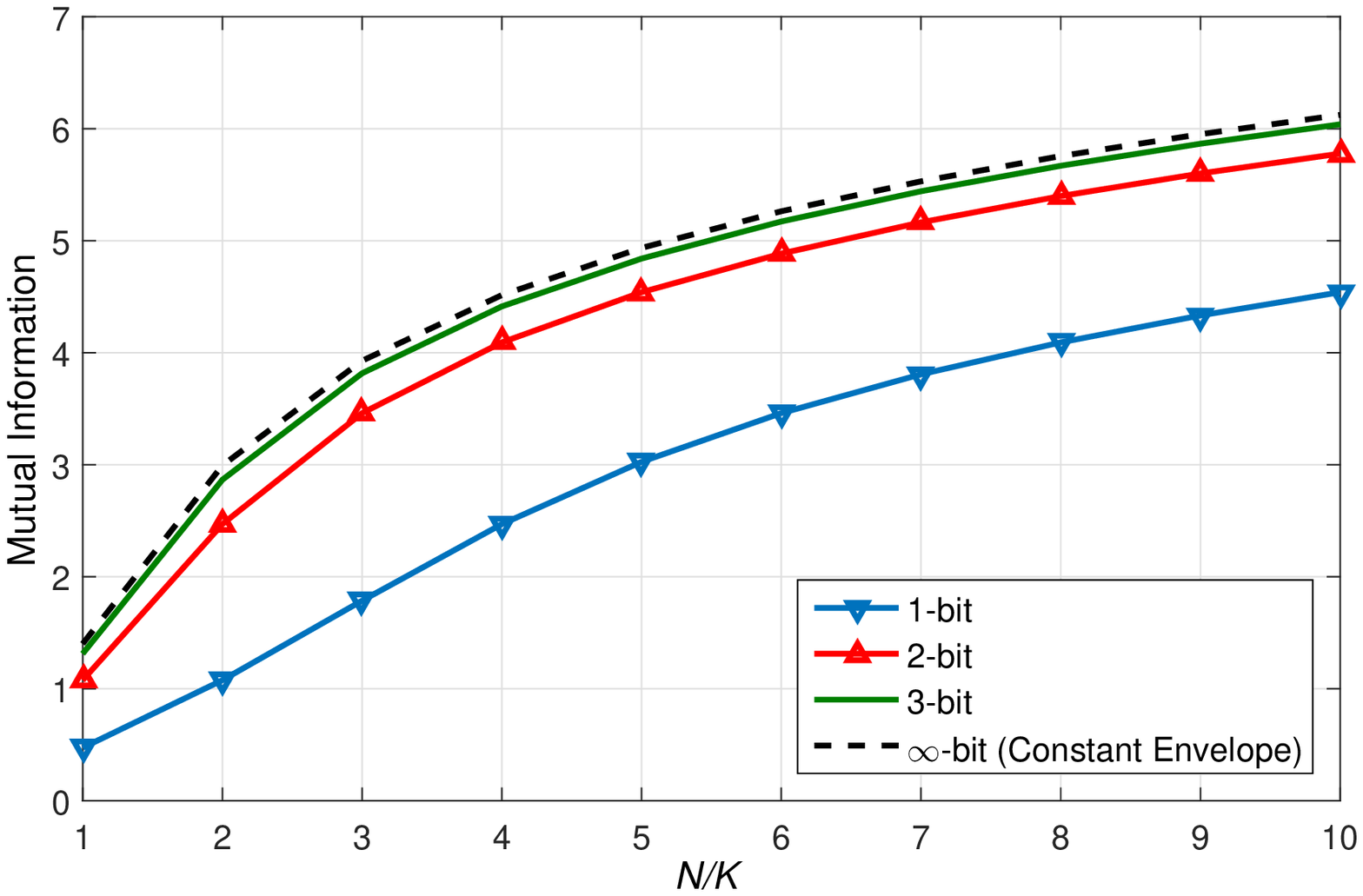}\vspace{-0.0cm}}
\caption{Theoretical mutual information, $C=\log_2(1+{\rm SINR}_{\rm opt})$ from (\ref{SINR_opt}), based on the state-evolution analysis for several phase DAC resolutions for $N/K=5$ (left) and SNR$=10$ (right). With only 3-bit DACs, ideal performance is nearly achieved.}
\label{sim4}
\end{figure}

In Fig.~\ref{sim4}, we compute the achievable rate obtained by the state evolution analysis in (\ref{SINR_opt}) as a function
of the SNR and for several phase DAC resolutions. The analytical curves predict that 3-bit is potentially sufficient to approach the ideal performance. 
 This result, however, might not be applicable for the very high SNR regime, as the non-rigorous state evolution analysis (or equivalently the replica method in \cite{Sedaghat_2018}) might fail in this regime. 

\subsection{Joint precoding and pulse shaping for uncorrelated Rayleigh flat fading channels}
\begin{figure}[htb]
\centering
\psfrag{xd}[c][c]{$\hat{s}$}
\centerline{\includegraphics[width=3.5in]{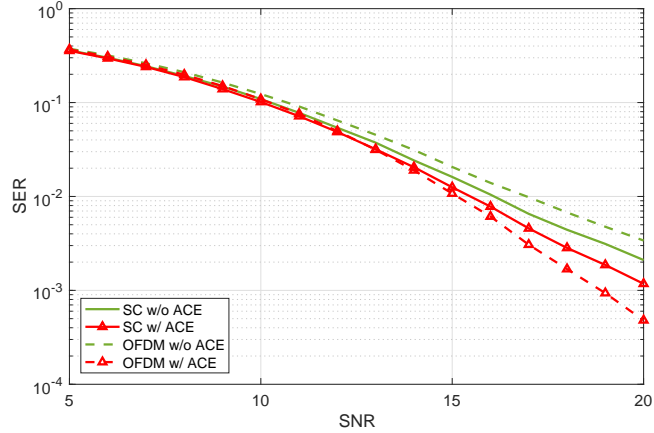}\vspace{-0.0cm}}
\caption{SC vs OFDM performance for $N=100$, $K=20$, $M=1024$, 16QAM constellation  and 2-bit phase DACs. Despite its common issue with the higher PAPR, OFDM with ACE performs slightly better than ACE. }
\label{sim5}
\end{figure}

The performance of the algorithm in terms of symbol error rate performance with OFDM and single carrier-OFDM (SC) is evaluated in Fig.~\ref{sim5} for flat channels with block Rayleigh-fading and IID coefficients. We consider a scenario with $N=100$ antennas, $K=20$ users, $M=1024$ FFT-size and 16QAM constellation. Additionally, 2-bit DACs with 2 samples/symbol are used.  Surprisingly, the performance of SC is slightly better than OFDM without ACE while an opposite observation can be made with the use of ACE. We explain this by the fact that the linear mixing through the FFT produces an IID-like matrix which helps to find better constellation points. Generally speaking, when using the proposed nonlinear method, the performance of OFDM is quite similar to SC and there is no penalty due to the higher Peak-to-Average Power Ratio (PAPR), as opposed to linear methods.

\begin{figure}[htb]
\centering
\psfrag{xd}[c][c]{$\hat{s}$}
\centerline{\includegraphics[width=3.5in]{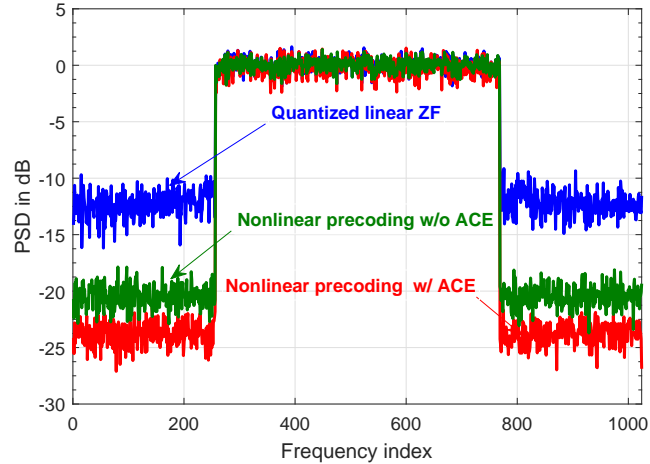}\vspace{-0.0cm}}
\caption{OFDM OTA spectral shaping performance of linear and nonlinear techniques as observed at the receivers locations without noise for $N=100$, $K=20$, $M=1024$, 16QAM constellation and 2-bit phase DACs.  More than 10dB OOB performance improvement with the proposed nonlinear precoding as compared to quantized linear ZF precoding. }
\label{sim6}
\end{figure}

In terms of OFDM spectral shaping performance, measured in the users' directions, Fig.~\ref{sim6}  shows for the same setting that adequate results can be still obtained even with just quadrature-phase DACs. The OTA power spectral density which is very close to the desired rectangular shape Adjacent Channel Leakage Ratio (ACLR) values of up to -25dBc can be achieved when combining nonlinear precoding with ACE.  Therefore, the proposed method can particularly cope with the scenarios of co-located users that utilize adjacent channels, which is quite surprising given the extremely low resolution of the DACs.

\begin{figure}[htb]
\centering
\psfrag{xd}[c][c]{$\hat{s}$}
\centerline{\includegraphics[width=3.5in]{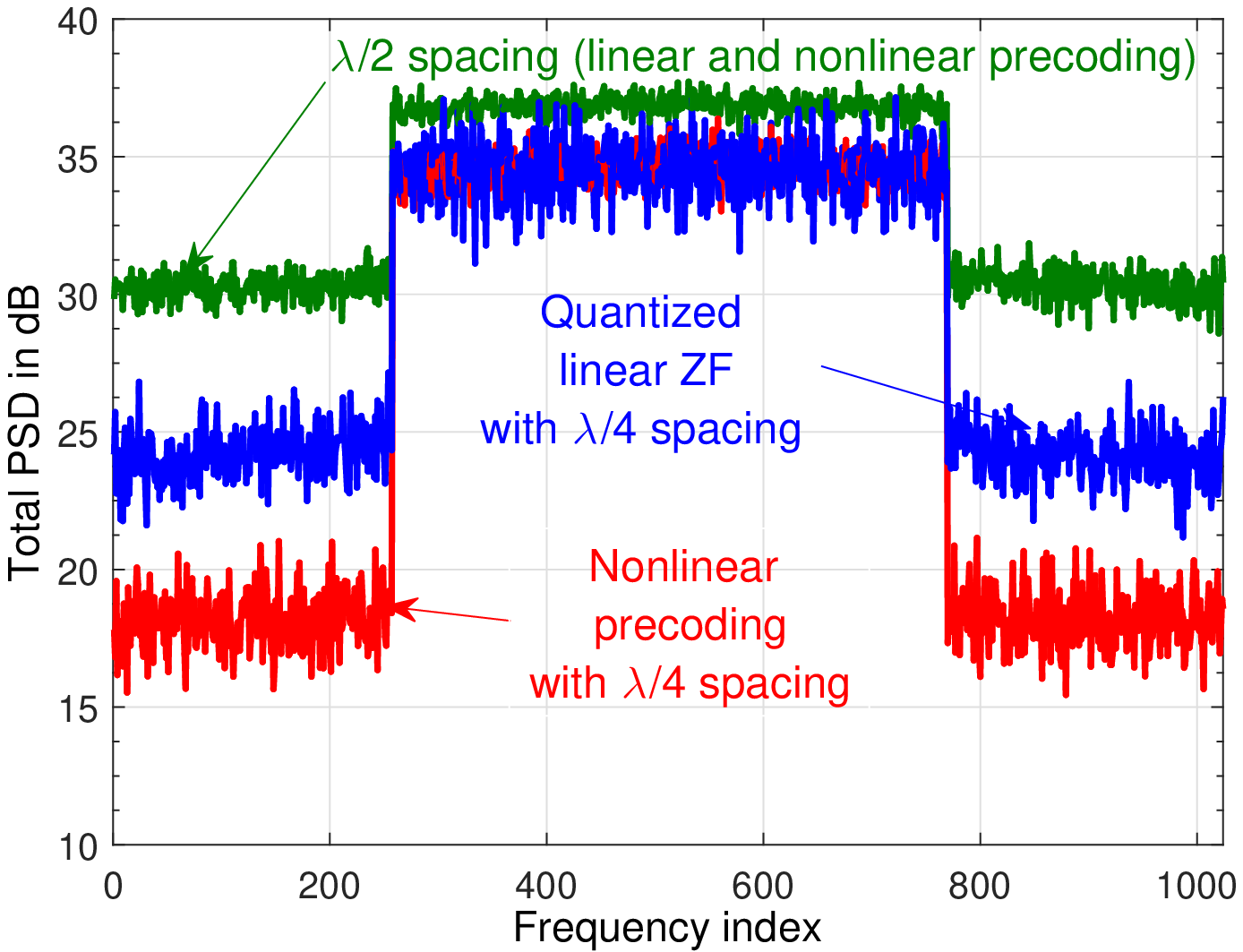}\vspace{-0.0cm}}
\caption{Total radiated power spectral density of linear and nonlinear techniques with $\lambda/2$ and $\lambda/4$ antenna-spacing for $N=10\times10$ uniform planar array, $K=20$, $M=1024$, QPSK constellation and 2-bit phase DACs. More than 10dB OOB performance improvement with quarter-wavelength spacing and nonlinear precoding.}
\label{sim61}
\end{figure}

Nevertheless, the performance in terms of power spectral density of the total radiated power defined in (\ref{radiated_power}) remains mediocre when critical half-wavelength sampling is used irrespectively of the precoding technique (linear or nonlinear) as shown in  Fig.~\ref{sim61}. Interestingly, decreasing the element spacing of a conventional uniform planar array to one-quarter wavelength can reduce the OOB total power by more than 10dB, particularly when the proposed nonlinear precoding method is used. The in-band unwanted radiations (in-band distortion and sidelobes) are also suppressed. Consequently, spatial oversampling can be considered as an enabler for low resolution signal conversion (much like temporal oversampling), in addition to being required for super-wideband antenna arrays \cite{Neto2006}.

\subsection{Joint precoding and pulse shaping for frequency selective channels}
Consider a uniform linear array (ULA) of $N$ hypothetical isotropic antennas with half-wavelength spacing at the center frequency $f_{\rm c}$. The  broadband array frequency response in the $M$-DFT domain assuming all the frequencies propagate with the same speed is \cite{Brady_2015}
\begin{equation}
\begin{aligned}
&\B{a}_{\rm ULA}(m,\varphi)\!=\!\!\left[1~ {\rm e}^{-{\rm j}{\pi} \cos(\varphi) (\frac{m}{TT_sf_{\rm c}}+1  )}\cdots {\rm e}^{-{\rm j}{\pi} \cos(\varphi) (N-1) (\frac{m}{TT_sf_{\rm c}}+1  ) } \right]^{\rm T},
\end{aligned}
\label{array_res}
\end{equation}
where $\varphi$ is the angle-of-departure (AoD), $T_s$ is the sampling interval and $m = -\frac{M}{2},\cdots,\frac{M}{2}-1  $ is the normalized frequency index.
We describe the wireless propagation channel 
by a sparse scattering model, where the $N$-dimensional
channel response $\B{h}_k[m]$ of user $k$ in the frequency domain  consists of the superposition
of $Q_k \ll N $ multi-path components. We express the resulting ray-based channel model\footnote{The model can be extended to include the filtering effects at the transmitter and receivers.} in the discrete frequency domain  as the following   
\begin{equation}
\begin{aligned}
\B{h}_k[m]= \sum_{\ell=1}^{Q_k} \alpha_{\ell,k} \B{a}_{\rm ULA}(m,\varphi_{\ell,k})  {\rm e}^{-{\rm j} 2 \pi m \frac{t_{\ell,k}}{M T_s}},
\end{aligned}
\end{equation}
where $\alpha_{\ell,k}$ are the path coefficients (including path phase and strength) and $\varphi_{\ell,k}$ and $t_{\ell,k}$ the associated AoD and time delay. 

\begin{figure}[htb]
\centering
\psfrag{xd}[c][c]{$\hat{s}$}
\centerline{\includegraphics[width=3.5in]{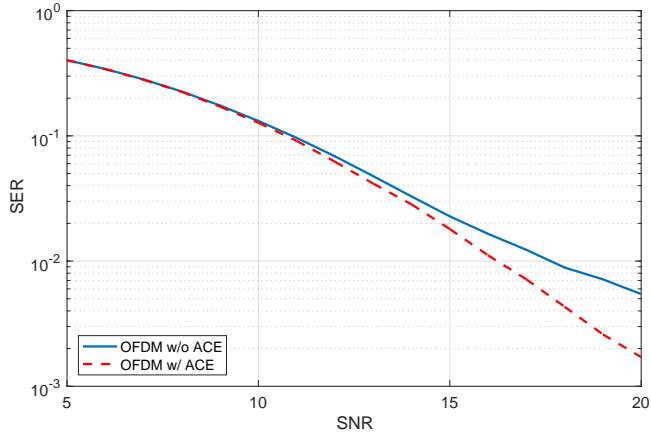} \vspace{-0.0cm}}
\caption{OFDM performance for frequency selective channles with $N=100$, $K=20$, $M=1024$, 16QAM constellation  and 2-bit phase DACs.}
\label{sim7}
\end{figure}

In the simulation that follows, we consider an OFDM wideband signal with sampling rate  $1/T_s=7$GHz, carrier frequency $f_c=60.5$GHz, and 16QAM modulation. As in the previous setting,  we use $K=20$ users, a uniform linear array of size $N=100$ with half-wavelength element spacing, $Q_k=10$ multi-path components for all users, and a coherence length of $M=1024$ samples. The center $M/2=512$ FFT points are filled with data. The scattering parameters $\alpha_{\ell,k}$ are  distributed as $\alpha_{\ell,k} \sim \mathcal{CN}(0,1/Q_k)$ and the path delays $t_{\ell,k}$  are selected from a uniform distribution in the interval $[0,50T_s]$.  The SER  performance is shown in Fig.~\ref{sim7} where the benefits of ACE are observed similarly to the frequency flat case. In addition, the performance degradation compared to frequency flat channels with IID coefficient in Fig.~\ref{sim5} is not significant.  

{
\subsection{Complexity and convergence}
 In terms of complexity, the number of computations per iteration in our algorithm is of the order $\mathcal{O}(NK)$ for each spatially processed transmission vector which is mainly due to the matrix-vector multiplications in step 5 and 6 of Algorithm~\ref{GAMP_alg}.  The per-iteration complexity is similar to gradient-based methods such as GEMM in \cite{Shao_2019}. However, GEMM is based on the SER criterion and involves more challenging computations due to the transcendental SER function. 


    The other benchmark method in Fig.~\ref{sim30},  P-BB \cite{Li_2020}, while being almost optimal, has an exponential worst-case complexity and is not scalable. If temporal processing is included for spectral shaping, such as OFDM or channel pre-equalization, then an additional per-iteration per-sample complexity of order $\mathcal{O}(N\log_2M )$ is added for the DFT and IDFT processing (c.f. (\ref{ym})). The number of iterations of the proposed algorithm generally does not increase with the size of the problem. In our simulation, it was observed that convergence is typically achieved within 20 iterations. All in all, the proposed method has a complexity per iteration that is of the same order as conventional linear methods. Therefore, the total complexity is in the range of 20-40 times the complexity of purely linear methods regardless of the channel and processing type.
 } 

\section{Conclusion}
The nonlinear precoding problem has been considered with low resolution phase DACs in the context of multi-user massive MIMO. A modified min-SUM GAMP algorithm with stochastic relaxation has been proposed to solve the corresponding minimization based on the MSE and the SER formulations.
We showed, theoretically as well as by means of simulations, that the MSE criteria, combined with active constellation extension, can provide nearly the same performance in terms of uncoded SER as the SER criteria.  We, therefore, concluded that the MSE  based design is indeed a very efficient and attractive criterion for dealing with the nonlinear precoding problem due to its mathematical tractability. 
We also performed a large system analysis by means of the state evolution approach for IID flat fading channels. 
An essential aspect of the analysis is the tight connection to the  GAMP algorithm allowing a more insightful way of studying the asymptotic behavior compared to previous work that relies on the replica method. An equivalent scalar channel model has been derived that mimics this behavior. 
 While the state-evolution analysis is mathematically non-rigorous in general, it can still provide appropriate performance predictions up to medium SNR values.
Finally, the joint nonlinear precoding and pulse shaping problem has been investigated for frequency selective channels based on the same GAMP algorithm, and we showed that the resulting OOB performance can be adequate even with 2-bit DACs per antenna, particularly when combined with spatial oversampling. It is worth mentioning that the proposed nonlinear precoding might additionally exhibit higher physical-layer security since the data is scrambled non-trivially in all directions other than the intended user \cite{Kalantari_2016}.   Taking into account the analog filtering after the DACs in combination with significant temporal oversampling would be an interesting extension of the work. 




%
\bibliographystyle{IEEEtran}     
\bibliography{references}{}
\end{document}